\documentstyle[12pt]{ioplppt}     

\jl{6}

\def\be{\begin{equation}}
\def\ee{\end{equation}}
\def\beq{\begin{eqnarray}}
\def\eeq{\end{eqnarray}}
\def\n{\nonumber}

\def\ts {\textstyle}

\begin{document}

\title{A causal model of radiating stellar collapse}

\author{
M. Govender\dag, S. D. Maharaj\dag\ and R. Maartens\ddag
}

\address{\dag\ Department of Mathematics and Applied 
Mathematics,
University of Natal, Durban 4041, South Africa
}

\address{\ddag\ School of Computer Science and Mathematics,
Portsmouth University, Portsmouth P01 2EG, Britain 
}

\begin{abstract}

We find a simple exact model of radiating stellar collapse, with
a shear-free and non-accelerating
interior matched to a Vaidya exterior.
The heat flux is subject to causal thermodynamics, leading
to self-consistent determination of the temperature $T$.
We solve for $T$ exactly when the mean
collision time $\tau_{\rm c}$ is constant, and
perturbatively in a more realistic case
of variable $\tau_{\rm c}$. 
Causal thermodynamics predicts temperature behaviour that
can differ significantly from the predictions of non-causal theory.
In particular, the causal theory gives a higher central
temperature and greater temperature gradient.

\end{abstract}

\pacs{0440, 4775, 9530, 9760}



\section{Introduction}

The problem of constructing physically realistic models for
radiating collapsing stars is of major importance to
relativistic astrophysics. Such models will necessarily involve
complicated numerical techniques for their efficient and reliable
solution. It is also useful however to construct simple exact models,
which are at least not physically unreasonable. This allows for
a more transparent analysis of the main physical effects at play,
and it can also serve as a useful check for numerical procedures.
In this spirit, we construct a simple exact model which 
generalizes a previous model \cite{g}
by incorporating the physically
desirable feature of causality into the heat transport process.

A non-rotating
spherically symmetric star that is radiating energy must be matched
to a Vaidya exterior spacetime. It is known \cite{s} that the
junction
conditions imply non-vanishing pressure at the boundary, due to the
presence of an energy flux. This is part of the reason for the
difficulty in finding explicit analytic forms for reasonable
interior solutions. Another difficulty arises from the physics of the
energy flux, which reduces to a heat flux in the absence of particle
flux. Physically consistent heat flux must be related to
the temperature gradient and four-acceleration via a thermodynamical
transport equation. Most models employ a relativistic Fourier
equation, but this is non-causal, since it leads to superluminal
wave-front velocities, and all its equilibrium states are
unstable \cite{hl}. It should be replaced by the causal transport
equation arising in the transient
thermodynamics of Israel and Stewart \cite{is} or the essentially
equivalent extended
thermodynamics \cite{jcl}. The aim of this paper is to incorporate
the relativistically consistent causal thermodynamics (as opposed
to the non-causal theories which are essentially not
consistent relativistic theories) in a simple model of a
non-rotating radiating star.

The application of causal thermodynamics to radiating
stellar collapse has recently been developed 
via physically detailed models \cite{m,dh,hdhmm},
whose solutions 
require complicated numerical integrations.
We aim instead for
a simple exact model as a complement to numerical
models with physical detail and complexity. We follow \cite{g}
in choosing a very simple Friedmann-like interior, i.e. a
fluid that collapses without shearing or accelerating.
The behaviour of the energy density and pressure
in these models was discussed in \cite{g}, and assessed to be
not physically unreasonable. Our model shares this feature.
The crucial difference is that \cite{g} employs the
Fourier equation to determine the stellar
temperature, introducing the undesirable feature of non-causal
heat transport.
We replace the non-causal transport equation used in \cite{g} by
a causal equation, in order to produce a more satisfactory model that
is constrained by causality.

As shown in \cite{m,dh,hdhmm}, the relaxational effects introduced
by the causal theory can have a significant and
in principle observable impact on the
temperature, rate of collapse and other properties.
Our simple exact model confirms this general point. We find that
although the collapse rate is unchanged, owing to the 
simple Friedmann-like
nature of the model, the causal temperature can differ significantly
from the non-causal prediction. In particular, the causal temperature
has greater central value and gradient.
This exact result is in agreement with the perturbative results of
\cite{hs}, which investigates the response of initially static stars 
to shear-free perturbations. 
The properties of our model are established in Section 4, where we 
find an exact solution for the temperature. In
Section 2 we briefly review the Friedmann-like stellar model, and
in Section 3 we discuss the causal heat transport equation. Finally,
concluding remarks and a perturbative solution for the temperature
are given in Section 5.

\section{The simple stellar model}

In isotropic and comoving coordinates, the non-rotating,
non-accelerating and shear-free interior
metric is given by \cite{g}
\beq
ds^2 &=& - dt^2 + A(t,r)^2
\left[dr^2 + r^2(d\theta^2 + \sin^2\theta
d{\phi}^2)\right] \,, \label{1} \\
A &=& \frac{M}{2b}\left[\frac{1 - b^2\lambda(t)}{1 -
r^2\lambda(t)}\right] u(t)^2 \,, \label{1a}
\eeq
where $u= (6t/M)^{1/3}$, $\lambda = a\exp u$, and $a$, $b$
and $M$ are constants. The fluid four-velocity is $u^\alpha=
\delta^\alpha{}_0$ (so that $t$ is comoving proper time), and
the four-acceleration $\dot{u}_\alpha\equiv u^\beta\nabla_\beta
u_\alpha$ vanishes. The fluid volume collapse rate is
\be
\Theta = 3{\dot{A} \over A}   \,,
\label{}
\ee
and since the shear vanishes, the collapse rates in the
radial and tangential directions both equal $\dot{A}/A$.

The heat flux (which is the total energy flux, since there is
no particle flux relative to $u^\alpha$) has the form
\[
q_\alpha =q(t,r)n_\alpha \,,
\]
where $n_\alpha$ is a unit radial vector,
so that $q$ is a
covariant scalar measure of the heat flux ($q^2=q^\alpha q_\alpha$).
The other dynamical covariant scalars are the energy density $\rho$
and isotropic pressure $p$.
The Einstein field equations
imply (using units with $c=1=8\pi G$) \cite{g}
\beq   \label{2}
\rho &=& \frac{12}{M^2 u^4}\left[\left\{\frac{2}{u} - \frac{(b^2 -
r^2)\lambda}{(1 - b^2\lambda)(1 - r^2\lambda)}\right\}^2 -
\frac{4b^2\lambda}{(1 - b^2\lambda)^2}\right] \,, \label{2a}  \\
p &=& \frac{4}{M^2u^4}\frac{(b^2 - r^2)\lambda}{(1 - b^2\lambda)(1 -
r^2\lambda)}\left[\frac{8}{u} + \frac{5}{1 - r^2\lambda} -
\frac{1}{1-
b^2\lambda} - 2\right] \n \\
&&{} + \frac{16}{M^2u^4}\frac{b^2\lambda}{(1 - b^2\lambda)^2} \,,
\label{2b} \\
q &=& \frac{16 br \lambda}{M^2 u^4(1 - b^2\lambda)
(1 - r^2\lambda)}\,.
\label{2c}
\eeq
Equations (\ref{1})--(\ref{2c}) comprise an exact
solution to the
Einstein field equations for the interior of the
radiating star.
This must match smoothly to the exterior Vaidya
spacetime across a comoving
time-like boundary, which we denote by $\Sigma$.
The Vaidya metric is the unique isotropic null-radiation solution,
and is given by
\be  \label{666}
ds^2 = - \left[ 1 - \frac{2m(v)}{ R }\right] dv^2 - 2dvd R +
 R ^2 \left(d\theta^2 + \sin^2\theta
d\phi^2 \right) \,,
\ee
where $m$ represents the
Newtonian mass of the gravitating body as measured by
an observer at infinity.

Matching of the metrics (\ref{1}) and (\ref{666}) gives the
junction conditions \cite{qq}
\beq
(R)_{\Sigma} &=& (rA)_{\Sigma} \,,\label{m1} \\
(p)_{\Sigma} &=& (q)_{\Sigma} \,, \label{m2} \\
\left[r(rA)'\right]_{\Sigma} &=&
\left[{\dot v}\left(R - 2m\right) + R\dot R
\right]_{\Sigma} \,, \label{m3} \\
m(v) &=& \left[{\ts{1\over2}}r^3A \dot{A}^2
 - r^2A' -{\ts{1\over2}}
r^3 A^{-1} A'^2\right]_{\Sigma} \,, \label{m4}
\eeq
where a prime denotes
$\partial/\partial r$
and the boundary is defined by $r_{\Sigma} = b = \mbox{constant}$.
It follows that the proper stellar radius (not given in \cite{g}) is
\[
r_{\rm p}(t)=\int_0^b A\,dr={M\over 2b\sqrt{\lambda}}
\mbox{artanh}\left(b\sqrt{\lambda}
\right)\left(1-b^2\lambda\right)u^2 \,,
\]
which is decreasing since $A$ is decreasing and $r$ is comoving.
The pressure at the centre follows from (\ref{2b}) as
\[
p_0 = {12b^2\lambda \over M^2u^4(1 - b^2\lambda)}\left[1+{8
\over 3u} + {1 \over {1 - b^2\lambda}}\right] \,,
\]
which becomes zero when
\be \label{w1}
1+{8 \over 3u}  + {1 \over {1 - b^2\lambda}} = 0 \,.
\ee
The time taken for the formation of the horizon is a solution of
\be \label{w2}
{2 \over u} + {{1 + b^2\lambda} \over {1 - b^2\lambda}} = 0 \,,
\ee
as established in \cite{w}. Equations (\ref{w1}) and (\ref{w2})
together with the strong energy conditions place the following
restrictions \cite{g}
\begin{eqnarray*}
0 \leq & ab^2& \leq (3 - \sqrt{8})e^{\sqrt{2}} \,, \\
- \infty \leq & u & \leq u_H \,,
\end{eqnarray*}
where $u_H= - \sqrt{2}$, the time of formation of the horizon, is a
solution of (\ref{w2}). The body starts collapsing at $u = -\infty$
with an infinite radius and zero density, and evolves to $u = u_H$.
Note that the results (\ref{w1}) and (\ref{w2}) are the
same as in the corresponding non-causal model of \cite{qq, w}.
The behaviour of the energy density $\rho$ and the pressure $p$
in the model (\ref{2a})--(\ref{2c}), for the line element (\ref{1}),
has been studied extensively in \cite{w}. The energy density is a
decreasing function in the interval $-\infty \leq u \leq u_H$. The
pressure gradient is negative in the early stages of collapse but at 
a
later epoch the pressure gradient becomes positive. It was 
established
in \cite{g} that the heat flow $q$ is a monotonically increasing
function of $r$ and $u$. Thus the behaviour of $\rho$, $p$ and
$q$ in the simple Friedmann-like solution is not
physically unreasonable.

\section{Causal heat transport}

We now apply the causal relativistic thermodynamics of Israel and
Stewart \cite{is} to give physical meaning to $q_\alpha$. The
causal transport equation in the absence of rotation and viscous
stress is (see \cite{mt} for the general case)
\be \label{3}
\tau h_\alpha{}^\beta
\dot{q}_\beta+q_\alpha = -\kappa \left(
h_\alpha{}^\beta\nabla_\beta T+T\dot{u}_\alpha\right)\,,
\ee
where $h_{\alpha\beta}=g_{\alpha\beta}+u_\alpha u_\beta$
projects into the comoving rest space,
$g_{\alpha\beta}$ is the metric, $T$ is the local equilibrium
temperature, $\kappa$ ($\geq0$) is the thermal conductivity and
$\tau$ ($\geq 0$)
is the relaxational time-scale which gives rise to the
causal and stable behaviour of the theory. The transport 
equation as well as expressions for
the thermodynamic coefficients $\kappa$ and
$\tau$ may be derived via relativistic kinetic theory using
the Grad 14-moment method \cite{is}.

The non-causal Fourier transport
equation has $\tau=0$, and reduces from an evolution equation to an
{\em algebraic} constraint on the heat flux. Intuitively, the
non-causal behaviour arises because
the heat flux is instantaneously brought to zero when
the temperature gradient and acceleration are `switched off'.

For the metric (\ref{1}), equation (\ref{3}) becomes
\be \label{4}
\tau\dot{q} + q =
-\frac{\kappa T'}{A}\,,
\ee
since $\dot{u}_\alpha=0$. The very simple form of the transport
equation (\ref{4}) is balanced by the complexity of the equations
(\ref{2a})--(\ref{2c}) and (\ref{m1})--(\ref{m4}).

For a physically reasonable model, we use the thermodynamic
coefficients for radiative transfer \cite{we,st,ui,fl}. In
other words, we are considering the situation where energy is
carried away from the stellar core by massless particles,
moving with long
mean free path through matter that is effectively in
hydrodynamic equilibrium, and that is dynamically dominant.
The thermal conductivity has the form
\be
\kappa =\gamma T^3{\tau}_{\rm c}\,,   \label{5}
\ee
where $\gamma$ ($\geq0$) is a constant and ${\tau}_{\rm c}$ is the
mean collision
time between the massless and massive particles.
A detailed analysis
in \cite{m} for the case of neutrinos generated by
thermal emission shows that
$\tau_{\rm c}\propto T^{-3/2}$ to a good approximation. 
Based on this, we will assume the power-law behaviour
\be
\tau_{\rm c} =\left({\alpha\over\gamma}\right) T^{-\sigma} \,,
\label{5b}\ee
where $\alpha$ ($\geq 0$) and $\sigma$ ($\geq 0$) are constants,
with $\sigma={3\over2}$ in the case of thermal neutrinos.
The mean collision
time decreases with growing temperature, as expected, except for
the special case $\sigma=0$, when it is constant.
This special case can only give a reasonable model for a limited
range of temperature.
Following \cite{m}, we assume that the velocity of thermal
dissipative signals is comparable to the adiabatic sound speed,
which is satisfied if the relaxation time is proportional to
the collision time, i.e.
\be
\tau =\left({\beta\gamma \over \alpha}\right) \tau_{\rm c} \,,
\label{5c}\ee
where $\beta$ ($\geq 0$) is a constant. We can think of
$\beta$ as the `causality' index, measuring the strength of
relaxational effects, with $\beta=0$ giving the non-causal case.
(A detailed discussion of the magnitude and relative importance
of stellar relaxation times is given in \cite{dh,hs}.)

Substituting (\ref{2c}) and the thermodynamic equations
(\ref{5})--(\ref{5c}) into the transport equation
(\ref{4}) leads to an equation for the temperature:
\be  \label{6}
\alpha T^{3 - \sigma}\frac{dT}{ds}  + {\beta}\left(\dot{f}+
{\ts{1\over2}}s\right)T^{-\sigma} + 1 = 0 \,,
\ee
where
\begin{eqnarray*}
s &=& \frac{4}{Mu^2(1 - r^2\lambda)} \,, \\
f(t) &=& -\ln\left[u^2(1-b^2\lambda) \right] \,.
\end{eqnarray*}
This is the fundamental equation for our simple causal model.
It is a radial differential equation for each instant of
proper time.
The solution of (\ref{6}) together with (\ref{2a})--(\ref{2c})
represents a complete model in which all thermo-hydrodynamical
quantities are known explicitly, and the model can be compared with
its non-causal counterpart, which is the special case $\beta=0$.

\section{Causal temperature}

The temperature equation (\ref{6}) is readily solved exactly
in the non-causal case
$\beta=0$, as in \cite{g}.
For the more satisfactory relativistic model $\beta>0$,
we have succeeded in integrating (\ref{6}) exactly
only when $\sigma = 0$,
i.e. when the mean collision time may be approximated as constant.
This can only be reasonable for a limited range of temperature,
but it is a useful solution for giving a qualitative idea
of the impact of causal relaxational effects, for 
checking numerical routines, and for
generating perturbative solutions
in the case of small $\sigma$ (see the following section).
Even when $\sigma=0$, the solution of (\ref{6}) is not simple. It
may be given in the form
\be  \label{7}
T^4 = -\frac{16 {\beta}}{\alpha}\left[\frac{1}{Mu^2(1 -
r^2\lambda)}\right]^2 - \frac{16}{\alpha}\left(\beta{\dot{f}}
+ 1\right) \left[\frac{1}{Mu^2(1 -
r^2\lambda)}\right] + F(t) \,,
\ee
where $F(t)$ is an integration function, which we can determine as
follows.
The effective surface temperature of a star is given by \cite{dh}
\be \label{8}
\left({T^4}\right)_{\Sigma} =
\left(\frac{1}{r^2A^2}\right)_{\Sigma}\left({L\over 4\pi\delta}
\right)\,,
\ee
where $\delta$ ($>0$)
is a constant and $L$ is the total luminosity at
infinity, which
 has the form
\be \label{9}
L = -\frac{dm}{dv} = \frac{2b^2\lambda}{(1 -
b^2\lambda)^2}\left[\frac{2}{u}  + \left(\frac{1 + b^2\lambda}{1 -
b^2\lambda}\right)\right]^2  \,.
\ee
We can evaluate (\ref{7}) at the comoving
boundary $r = b$ to find $F(t)$ with
the help of (\ref{8}) and (\ref{9}). This yields
\begin{eqnarray}
F(t) &=& \frac{16 {\beta}}{\alpha}\left[\frac{1}{Mu^2(1 -
b^2\lambda)}\right]^2 + \frac{16}{\alpha}\left(
{\beta}{\dot{f}} + 1\right)
\left[\frac{1}{Mu^2(1 -
b^2\lambda)}\right] \n \\
&&{} +  \frac{2b^2\lambda}{\pi \delta M^2u^4(1 -
b^2\lambda)^2}\left[\frac{2}{u}  + \left(\frac{1 + b^2\lambda}{1 -
b^2\lambda}\right)\right]^2 \,.
\label{10}
\end{eqnarray}
Finally, the temperature has the
explicit exact form
\begin{eqnarray}  \label{11}
T^4 &=& \frac{16 \lambda(b^2 - r^2)}{\alpha
M(1 - b^2\lambda)(1 - r^2\lambda)u^2}\left\{\beta \frac{[2 - (b^2 +
r^2)\lambda]}{Mu^2(1 - b^2
\lambda)(1 - r^2\lambda)}  + {\beta}{\dot{f}} + 1
\right\}  \n \\
&&{}+  \frac{2b^2\lambda}{\pi \delta M^2u^4(1 -
b^2\lambda)^2}\left[\frac{2}{u}  + \left(\frac{1 + b^2\lambda}{1 -
b^2\lambda}\right)\right]^2 \,.
\end{eqnarray}
The temporal and spatial dependence are specified fully, and together
with the expressions (\ref{2a})--(\ref{2c}), this represents a
complete exact model for causal radiating stellar collapse.

The non-causal temperature $\tilde{T}$
is obtained by setting $\beta = 0$ in
(\ref{11}). It is interesting to note that the non-causal and causal
temperatures coincide at the surface $r = b$ of the radiating star:
\be
T(t,b)=\tilde{T}(t,b) \,.
\label{t1}\ee
However, it is clear from (\ref{11}) that at all interior points,
the causal and non-causal temperatures differ.
In particular, we observe that the
causal temperature is greater than the non-causal temperature at the
centre of the star: 
\be
T(t,0)>\tilde{T}(t,0) \,.
\label{t2}\ee
For small values of $\beta$ the temperature
profile is similar to that of the non-causal theory, but
as
$\beta$ is increased, i.e. as relaxational effects
grow, it is clear from (\ref{11}) that
the temperature profile can deviate substantially from that of 
the non-causal theory.

It follows from (\ref{t1}) and (\ref{t2}) that the causal
temperature has a greater average gradient. In fact, 
we can show that the gradient
is greater at each $r$ even when $\sigma>0$. 
From the transport equation 
(\ref{4}) for non-accelerating collapse, we note that
\[
\kappa(T)T' - \kappa(\tilde{T}){\tilde T}' = -(A\tau)
{\dot q} \,.
\]
Using the radiative form (\ref{5}) for $\kappa$ and the power-law
generalization (\ref{5b}) of neutrino transport for $\tau_{\rm c}$, 
this becomes
\be
\left(T^{4-\sigma}\right)'-\left(\tilde{T}^{4-\sigma}\right)'
=-\left({4-\sigma\over\alpha}\right)(A\tau)\dot{q} \,.
\label{tgrad}\ee
(Note that this is 
independent of the particular form 
(\ref{5c}) for $\tau$.)
Hence the
relative radial gradient of the temperatures is governed 
by ${\dot q}$ and by the collision-time index $\sigma$.
It can be shown \cite{g}
from (\ref{2c}) that ${\dot{q}} > 0$. 
Since $\sigma={3\over2}$ for thermal neutrino 
transport \cite{m},
we are justified in assuming that $\sigma-4<0$.
It follows from (\ref{tgrad})
that the causal temperature gradient is everywhere
greater than that of the non-causal
temperature,
and that the difference grows with increasing $\tau$.
(Note that this conclusion still holds in the case $\sigma>4$.)

This particular exact result in a simple model is in agreement with 
the general result of \cite{hs}, i.e. that
for shear-free perturbations, the causal temperature gradient 
is greater. Our simple exact model
provides non-perturbative support
for the perturbative result. As pointed out in \cite{hs}, the fact
that the causal
temperature gradient is greater
for a given luminosity, is consistent with the
numerical results of \cite{dh}, which show that for a given
temperature gradient, the causal theory leads to lower luminosity.
Thus the numerical non-perturbative, the perturbative, and the exact
non-perturbative results are consistent.

\section{Concluding remarks}

By combining a simple stellar solution with 
physically consistent causal thermodynamics, we have been able to
develop an exact model of radiating stellar collapse, in which
it is straightforward to identify the relaxational effects of
the causal theory without resort to highly complicated numerical
methods. This should be seen as a complement to the physically
more realistic and detailed models and their numerical integration
\cite{m,dh} and perturbative solution \cite{hdhmm,hs}. 
We showed that the causal temperature decreases radially
outward more steeply than the non-causal temperature, regardless
of the particular form of the relaxation time
$\tau$, and in agreement with the independent perturbative results 
of \cite{hs}. 
For the case where $\tau\propto\tau_{\rm c}$, i.e. 
where (\ref{5c}) holds,
and assuming constant collision 
time, i.e. $\sigma=0$ in (\ref{5b}), we 
found the exact solution (\ref{11}) of the temperature differential
equation (\ref{6}), using the luminosity 
to evaluate the constant of integration. This exact solution
predicts that the causal temperature coincides with the
non-causal temperature at the surface, but differs at all interior
points during the collapse, and is greater at the centre.

Our results confirm in a highly simplified but also
transparent form, the overall conclusion of the more
detailed models, i.e. that causal thermodynamics can introduce
fundamentally different behaviour, with potentially significant
implications in astrophysics. In this sense, our results
are an additional motivation for
further study of causal relativistic
models of stellar collapse. 

For a more realistic model, the mean collision time will
grow with decreasing temperature, i.e. we have $\sigma>0$ in
the temperature equation
(\ref{6}). 
Using the $\sigma=0$ solution (\ref{11}), we can solve
perturbatively in the case of small temperature
parameter $\sigma$.
Let
\[
T = T_0 + \sigma T_1 +O(\sigma^2) \,,
\]
where $T_0$ is the zero order ($\sigma=0$) solution
(\ref{11}).
Substituting into (\ref{6}) and linearizing,
we obtain the following ordinary differential equation in
$T_1$:
\[
\alpha \frac{dT_1}{ds} + (3 -\sigma)
\frac{T_1}{T_0}\frac{dT_0}{ds} = 0 \,,
\]
which easily integrates to give
\[
T_1 = \varphi {T_0}^{\sigma-3} \,,
\]
where $\varphi$ ($\geq0$) is a constant. Hence we may write
\be
T = T_0\left[1 + \sigma\varphi {T_0}^{\sigma-4}\right] \,,
\label{12}\ee
to first order in $\sigma$.
The effect of $\sigma$ is to increase $T$, i.e. to
retard the cooling due to heat 
transport,
with the correction being greater in cooler regions (near the 
surface)
and less in hotter regions (near the centre) of the
collapsing star, since $\sigma\ll 1$ ensures that $\sigma-3<0$.

\[ \]

\end{document}